\documentstyle[aps,prl]{revtex}
\begin{document}
\input epsf.sty
\twocolumn[\hsize\textwidth\columnwidth\hsize\csname %
@twocolumnfalse\endcsname
\draft
\widetext
\title{Charge Ordering and Polaron Formation in the Magnetoresistive 
Oxide La$_{0.7} $Ca$_{0.3}$MnO$_{3}$}
\author{C. P. Adams,$^{1,2}$ J. W. Lynn,$^{1,2}$ Y. M. Mukovskii,$^3$ 
A. A. Arsenov,$^3$ and D. A. Shulyatev$^3$}
\address{$^1$NIST Center for Neutron Research, National Institute\\
of Standards and Technology, Gaithersburg, MD 20899-8562}
\address{$^2$Department of Physics, University of Maryland, \\
College Park, MD 20742}
\address{$^3$Moscow Steel and Alloys Institute, Moscow 117936, Russia}
\date{\today}
\maketitle

\begin{abstract}
Neutron scattering has been used to study the nature of the ferromagnetic
transition in a single crystal of the perovskite 
La$_{0.7}$Ca$_{0.3}$MnO$_{3}$.  Diffuse scattering from lattice 
polarons develops as the
Curie temperature is approached from below, along with short range polaron
correlations that are consistent with stripe formation. Both the 
scattering due to the polaron correlations and the anomalous quasielastic
component in the magnetic fluctuation spectrum maximize very close to 
$T_{C}$, in a manner remarkably similar to the resistivity, indicating 
that they have a common origin.
\end{abstract}


\phantom{.}
]
\narrowtext

One of the simplifying features of conventional isotropic ferromagnets such
as Fe, Co, Ni, and EuO is the isolation of the spin system from the
underlying lattice. This made these materials ideal candidates to
investigate the low temperature magnetic excitations, critical dynamics,
scaling behavior, and nature of second-order phase transitions. This
situation contrasts dramatically with the colossal magnetoresistive (CMR)
oxides, exemplified
by the ferromagnetic doping regime of La$_{1-x}$Ca$_{x}$%
MnO$_{3}$ (0.15\mbox{$\, < $}$\, x \, $\mbox{$< \, $}0.5)%
\cite{GenRefs}. The transport in these perovskites
is directly connected to the magnetic system through the double exchange
mechanism, while the Jahn-Teller distortion of the Mn$^{3+}$O$_{6}$ couples
the magnetic and lattice systems, resulting in the electronic, lattice, and
magnetic degrees of freedom being intimately intertwined \cite
{GenRefs2,Theory}. The ground state spin dynamics has a number of unusual
properties such as large linewidths and anomalous dispersion \cite
{hwang98,doloc98}, while the combined metal-insulator and ferromagnetic
transition has been found to be quite different from conventional isotropic
ferromagnets \cite{lynn96,fernandez98,dolocJAP}. In particular, the spin wave
stiffness does not collapse as $T\rightarrow T_{C}$, but instead a
quasielastic diffusive component develops in the excitation spectrum. Above $%
T_{C}$, the conductivity is characterized by hopping that is believed to be
associated with polarons \cite{GenRefs,lynn96,billinge96,deteresa97,polaron},
and recently direct evidence for the formation of lattice polarons has been
observed in single crystals of the layered manganite La$_{1.2}$Sr$_{1.8}$Mn$%
_{2}$O$_{7}$ \cite{doloc99}, and in the cubic, half-doped (Nd$_{0.12}$Sm$%
_{0.88}$)$_{0.52}$Sr$_{0.48}$MnO$_{3}$ material \cite{shimomura99}. However,
the relationship between the ferromagnetic transition and the polaron
formation associated with the metal-insulator transition has remained a
mystery. In the present work we observe the formation of lattice polarons in
an optimally-doped cubic CMR material, and find clear
evidence for polaron ordering in the paramagnetic phase that is consistent
with some models of stripe formation. More importantly, the temperature 
dependence of this
polaron intensity develops simultaneously with the quasielastic spin
fluctuation scattering, directly connecting these two phenomena 
with the resistivity.

The sample is a $0.7$ g single crystal grown by the floating zone
technique \cite{Mukovskii}, 
with a single-peaked mosaic less than 0.25${^{\circ }}$. At this composition
the crystal structure is orthorhombic, but the distortion is small and the
domains are equally populated. Therefore for simplicity we will employ cubic
notation ($a=3.867$~{\AA} at room temperature), where nearest-neighbor
manganese atoms are along the [100]-type directions. Most measurements were
taken in the $(hk0)$ plane on the BT-2 and BT-9 triple-axis spectrometers at
NIST, using a variety of incident energies (13.7, 14.7, 30.5, and 50 meV)
and collimations. For the unpolarized neutron measurements the monochromator
and analyzer crystals were pyrolytic graphite (PG), while for polarized beam
measurements Heusler alloy polarizers were employed. PG filters were used
when appropriate to suppress higher-order wavelength contaminations. 
Statistical uncertainties quoted represent one standard deviation.

A typical example of the magnetic fluctuation spectrum observed below the
Curie temperature ($T_{C}=257$~K) is shown in Fig. 1(a). A $q$-independent
nuclear incoherent scattering of 2.6 cts/min and a flat background of 1.6
cts/min have been subtracted from these data. At this reduced wave vector of
(0.09,0,0) well defined spin wave excitations are observed in neutron energy
gain ($E$%
\mbox{$<$}%
0) and energy loss ($E$%
\mbox{$>$}%
0), and the solid curve is a least-squares fit of the spin wave cross
section, along with the quasielastic component centered at $E=0$, convoluted
with the instrumental resolution. Data taken at a series of $q$'s reveal
that the spin waves obey the usual quadratic dispersion law $E_{SW}=\Delta
+D(T)q^{2}$, with a
\begin{figure}
\hspace{-6mm}
\centerline{\epsfxsize=3.7in\epsfbox{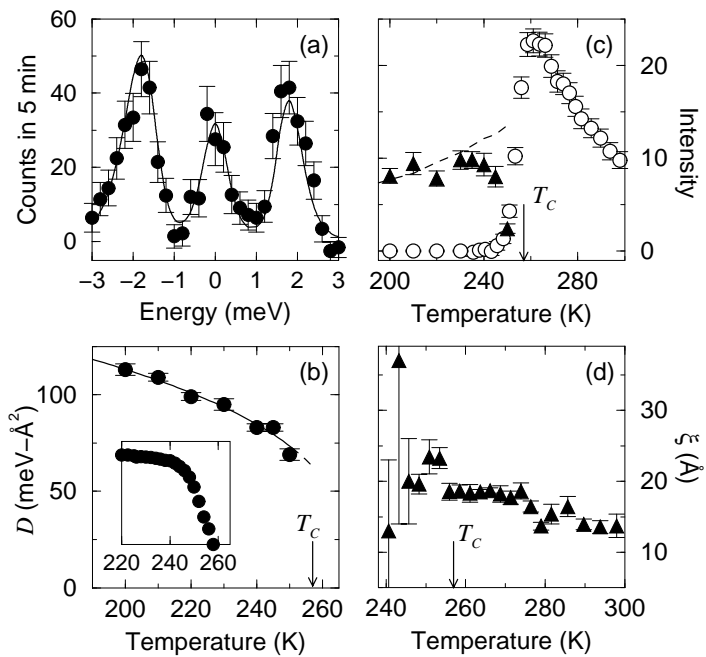}}
\caption[four panel]{(a) Constant-$Q$ scan taken at 
$T=240$~K with reduced wave vector $(0.09, 0, 0)$, showing the spin waves 
and quasielastic central peak. (b) Spin wave stiffness $D(T)$, which does not
appear to fall to zero at $T_C=257(1)$~K, as indicated by the
relative magnetization (inset).
Extinction of the (100) Bragg peak causes an unphysical flattening
of the magnetization below $\sim 240$~K.
(c) Intensity of spin waves (triangles) and the central peak (circles)
versus temperature.  The spin wave intensity follows 
Bose-Einstein statistics (dashed line) until the central peak
develops and eventually dominates the spectrum.
(d)~Correlation length associated with the
central peak, which is weakly temperature 
dependent and does not diverge at the ferromagnetic 
transition.}
\end{figure}
negligible spin wave gap $\Delta $, indicative of an isotropic 
ferromagnet, as has been found for all these CMR ferromagnets
\cite{LynnReview}. The temperature dependence of the spin stiffness
coefficient $D(T)$ obtained from these fits is shown in Fig. 1(b), which can
be compared with the magnetization determined from the (100) Bragg peak
(inset). A well defined Curie temperature of $T_{C}=257(1)$~K is obtained,
and there does not appear to be any significant distribution of $T_{C}$'s in
the sample. The spin wave stiffness does not seem to collapse at 
$T_{C}$, but instead
we see the development of the quasielastic component as shown in Fig. 1(a).
The energy width of this scattering is quadratic in wave vector, with a spin
diffusion constant $\Lambda =15(7)$~meV-\AA$^{2}$ (in this symmetry 
direction).  The temperature evolution of this component is shown in 
Fig. 1(c), where we
see that it attains its maximum intensity close to $T_{C}$.
The spin wave intensities, on the
other hand, show a decrease as the central component develops, rather than
the usual increase of the Bose-Einstein population factor. Finally, Fig.
1(d) shows the correlation length obtained from the $q$-dependence of the
quasielastic scattering. Below $T_{C}$ the quasielastic scattering is well
separated from the spin wave contribution with this relatively good energy
resolution (0.15~meV FWHM), while above $T_{C}$ all the scattering is
quasielastic. We find a length scale $\sim 16$~{\AA} that is only weakly
temperature dependent. This central peak, with the associated short length
scale, has been interpreted as the spin component of polarons \cite
{lynn96,deteresa97}. All these detailed results obtained on this high quality
single crystal are in good overall agreement with previous measurements on
polycrystalline samples \cite{lynn96}, as well as single crystal results on
related $x=1/3$ doped compounds \cite{fernandez98}.

The scattering from a lattice polaron arises from the
structural distortion that surrounds a carrier, and traps it. Individual
polarons generate diffuse (Huang) scattering around the fundamental
Bragg peaks, and we observe such diffuse scattering in the present La$_{0.7}$%
Ca$_{0.3}$MnO$_{3}$ crystal, which is similar to the scattering from polaron
distortions recently reported in the layered manganite La$_{1.2}$Sr$_{1.8}$Mn%
$_{2}$O$_{7}$ \cite{doloc99}, and in the half-doped cubic (Nd$_{0.12}$Sm$%
_{0.88}$)$_{0.52}$Sr$_{0.48}$MnO$_{3}$ material \cite{shimomura99}. The
temperature dependence of this diffuse scattering is shown in Fig. 2 for a
wave vector of $(1.85,2,0)$. The signal increases rapidly as the Curie
temperature is approached from below as the polarons form, while above $%
T_{C} $ we observe only a weak temperature dependence. This suggests that
the number of polarons increases rapidly as $T\rightarrow T_{C}$, while
above $T_{C}$ the number is roughly constant.

In addition to the diffuse polaron scattering, we observe well-developed
polaron-polaron correlations, which give rise to satellite peaks such as
those shown in Fig. 3. This is a scan of the elastic scattering measured
around the (4,0,0) nuclear Bragg peak. In the paramagnetic state (280 K) we
see two satellite peaks, indicating an ordering wave vector $(\frac{1}{4}%
,\frac{1}{4},0)$ and equivalent directions. The background scattering 
at 220 K is also shown, and
has a gentle variation which is due to the proximity of an aluminum powder
diffraction line from the sample holder;
the lower part of the figure shows
the subtraction of the 
\vskip -3mm
\begin{figure}
\hspace{12mm}
\centerline{\epsfxsize=2.9in\epsfbox{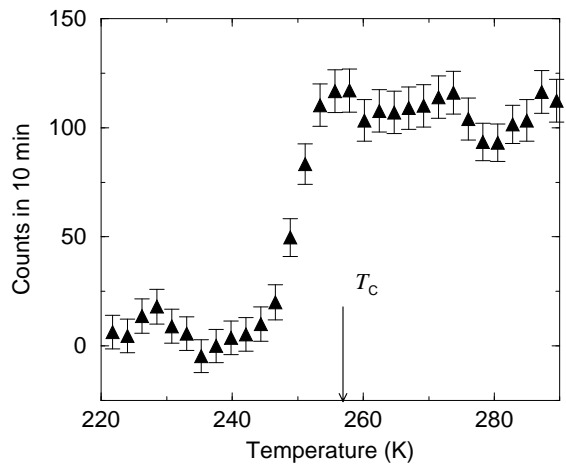}}
\caption[diffuse]{The temperature
dependence of the diffuse polaron scattering measured at $(1.85,2,0)$.}
\end{figure}
\begin{figure}
\hspace{-6mm}
\centerline{\epsfxsize=2.5in\epsfbox{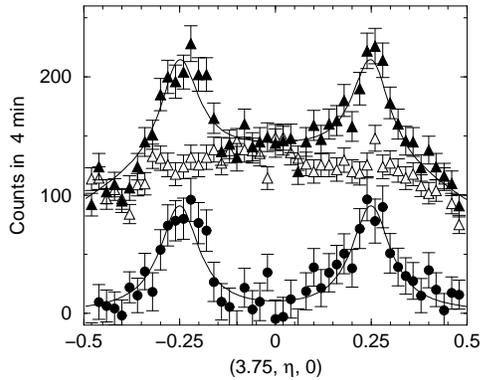}}
\vskip -1mm
\caption[chg order]{Transverse elastic scans through $(3.75,0,0)$ taken at 
220~K and 280~K (open and filled triangles respectively), showing the clear 
development of polaron peaks at $(\frac{1}{4},\frac{1}{4},0)$.
The subtracted data are shown as full circles. The solid curves 
are fits to Lorentzian peaks.}
\end{figure}
\vskip -2mm
two data sets to obtain the net scattering. These
polaron peaks have also been observed around other Bragg peaks such as the
(2,0,0), (3,0,0), (5,0,0), (3,1,0), (4,2,0), and (2,2,0), while the
intensity around the (0,0,0), (1,0,0), (1,1,0), and (2,1,0) was below our
detection limit. This trend for the intensity to increase with increasing $Q$
suggests that the dominant contribution to this scattering is from the
lattice. This has been directly confirmed by polarized beam measurements,
which show that the scattering in these peaks is dominated by the
non-spin-flip (lattice) component.

A familiar model that can explain both
the ordering wave vector of $(\frac{1}{4},%
\frac{1}{4},0)$ for the lattice polaron scattering as well
as the overall behavior of
the observed intensities of the short-range correlation peaks is 
the CE model \cite{GenRefs,GenRefs2} developed for the half-doped
case.  The CE model has an orbitally ordered Jahn-Teller lattice
with charge stripes in the [110] direction, and an antiferromagnetic
ground state.  Such ordering has been observed for $x \geq \frac{1}{2}$,
with the lattice and antiferromagnetic ordering at low
temperatures \cite{mori98}.  This contrasts with the present 
ferromagnetic system, where the orbital/charge correlations are
short range in nature, occur only {\em above} $T_C$, and
are not accompanied by antiferromagnetic correlations.
The lattice part of the CE model would of course need to be 
modified to accommodate
the smaller doping level, but with the same wave vector;  
so far there is no
apparent change in the ordering wave vector with doping.
In particular, in the half-doped material the diffuse signal
overwhelmed any peaks in the scattering, making it difficult to identify
correlations \cite{shimomura99}, but the weak structure observed suggested
correlations with a wave vector of magnitude $\sim $0.3-0.35 and in the
[110] direction. In the $x=0.4$ bilayer system clear peaks were observed, but
the wave vector is in the [100] direction, and of (incommensurate) magnitude
0.30 \cite{doloc99}. One similarity, however, is that the charge peaks for
La$_{0.7}$Ca$_{0.3}$MnO$_{3}$ and the bilayer both have a substantial
intrinsic width, indicating that the polaron ordering is short range in
nature. The width of the peaks is only weakly temperature dependent, and
yields a correlation length of $\sim 10(2)$~{\AA} (Fig. 3), which is the same
basic length scale observed for the quasielastic magnetic scattering 
[Fig. 1(d)].
The polaron peaks in both materials are also elastic, indicating that the
polarons are static on a time scale of 1~ps. The polarons are surely
hopping, though, and one of the interesting avenues to explore
experimentally will be to investigate the nature of these peaks with much
higher energy resolution, and in particular to determine if the observed $q$%
-widths are related to dynamics of the polarons instead of static short
range order.

Stripe formation has been observed in the related cuprates and nickelates,
but the intrinsic magnetism in those systems is always antiferromagnetic in
nature. This gives rise to separate satellite peaks associated with
the charge
and spin order \cite{tranquada94,lee97}. We have found no evidence for
separate magnetic satellites in La$_{0.7}$Ca$_{0.3}$MnO$_{3}$, which is not
surprising in this ferromagnetic system where the spin stripes 
(or spin polaron correlations) would be expected to give a contribution 
at the same satellite positions as the
charge satellites. It might then seem surprising that we observe no
significant magnetic component to the satellite peaks. However, the nature
of the two types of scattering is quite different. The lattice scattering
originates from well-formed (static, on this time scale) Jahn-Teller
distorted MnO$_{6}$ octahedra, which then have a correlation
length of $\sim 10$~{\AA}. The spin stripes would have the same short 
correlation length, but the magnetic scattering itself has a short 
correlation range of only $\sim 15$~{\AA} [Fig. 1(d)]. Model
calculations then show that the combination of the two short correlation
ranges renders the spin stripe scattering too
weak to be observed at any of the satellite positions. We conclude then that
the scattering of the lattice component of the polarons occurs around 
the high-$Q$ fundamental
Bragg peaks, along with the broad satellite peaks associated with
polaron-polaron correlations, while the dominant magnetic contribution
occurs around the low-$Q$ fundamental Bragg reflections, in the form of the
quasielastic scattering as we now discuss.

The temperature dependence of the intensity of the satellite peak at 
(3.75,0.25,0) is shown in Fig. 4. We see that the scattering begins to
develop in this sample $\sim 30$~K below $T_{C}$, rapidly develops as $%
T\rightarrow T_{C}$, and peaks just above the ordering
temperature. This behavior is very similar to the temperature
dependence of the resistivity, which is shown as the dashed curve
in Fig. 4 (scaled to the peak in the scattering).  The intensity of the 
quasielastic component
of the spin fluctuation spectrum [measured at (1.03,0,0)] has been similarly
scaled and is also shown in
the figure. We see that the temperature dependence of the two types of
scattering through the ferromagnetic phase transition is virtually 
identical to the resistivity, indicating that they all have a common origin.
\begin{figure}[tbp]
\hspace{-5mm}
\centerline{\epsfxsize=3.2in\epsfbox{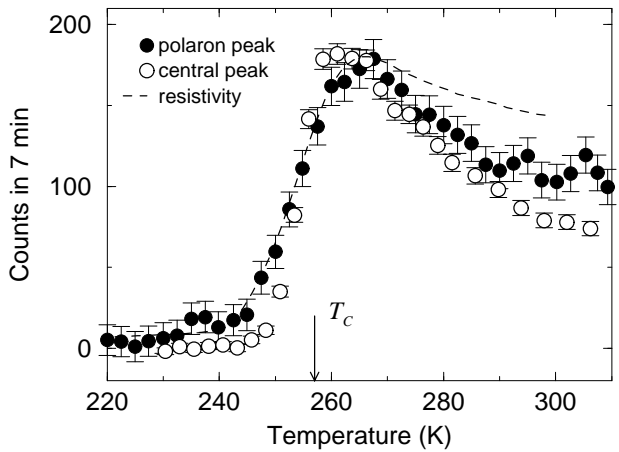}}
\caption[Tdep of chg/ cen peak]{Temperature dependence of the intensity 
of the polaron peak at $(3.75,0.25,0)$, compared to the central peak 
scattering at a wave vector of $(1.03,0,0)$. The data have been 
scaled so the peak heights match; 
the count rate describes the charge peak. The dashed curve is
the resistivity, scaled in the same manner. The similarity of the data 
indicate a common physical origin.}
\end{figure}

The above experimental results reveal that both the spin and charge
correlations associated with the polarons in La$_{0.7}$Ca$_{0.3}$MnO$_{3}$
appear together, and have a very similar spatial and temperature dependence.
The metal-insulator crossover in the conductivity also occurs close to 
$T_{C}$. This coincidence may
explain the amplified magnetoresistive effects, as well as the absence of
conventional magnetic critical behavior, both in the
Ca-doped system as well as in other materials\cite{fernandez98}. This
behavior is not universal, however. In the higher $T_C$ Sr and Ba 
systems, for example,
the polaron formation\cite{doloc98,dolocJAP,LaBa} and conductivity crossover%
\cite{Srconduct} can occur at temperatures substantially higher than the
Curie point, reducing the magnetoresistive effects and rendering the spin
dynamics more conventional. The physical quantities that control the
characteristic temperature for polaron formation, including the role that
metallurgical defects play in their formation, still remain to be elucidated.

The work at UM was supported by the NSF, DMR 97-01339, and NSF-MRSEC, DMR
00-80008. The work in Moscow was supported by ISTC grant \#636.

\end{document}